\documentclass[12pt,preprint]{aastex} 
\usepackage{graphicx}
\usepackage{epstopdf}

\tighten
\newcommand{\asec}{\hbox to 1pt{}\rlap{$^{\prime\prime}$}.\hbox to 2pt{}}
\newcommand{\amin}{\hbox to 1pt{}\rlap{$^{\prime}$}.\hbox to 2pt{}}
\newcommand{\atlas}{ATLAS$^{\rm 3D}$\ }
\newcommand{\ser}{S\' ersic\ }

\shortauthors{Postman et al.}
\shorttitle{The Core of A2261-BCG}

\begin{document}

\title{Cores and the Kinematics of Early-Type Galaxies}

\author{Tod R. Lauer\altaffilmark{1}}

\altaffiltext{1}{National Optical Astronomy Observatory,\footnote{The
National Optical Astronomy Observatory is operated by AURA, Inc.,
under cooperative agreement with the National Science Foundation.}
P.O. Box 26732, Tucson, AZ 85726, USA}

\begin{abstract}
I have combined the Emsellem et al.\ \atlas rotation measures of a
large sample of early-type galaxies with {\it HST}-based classifications
of their central structure to characterize the rotation velocities of
galaxies with cores.  ``Core galaxies'' 
rotate slowly, while ``power-law galaxies'' (galaxies that lack
cores) rotate rapidly, confirming the analysis of Faber et al.
Significantly, the amplitude of rotation sharply
discriminates between the two types in the $-19>M_V>-22$ domain
over which the two types coexist.  The slow rotation
in the small set of core galaxies with $M_V>-20,$ in particular,
brings them into concordance with the more massive core galaxies.
The \atlas ``fast-rotating'' and ``slow-rotating'' early-type
galaxies are essentially the same as power-law and core galaxies,
respectively, or the Kormendy \& Bender two families of elliptical
galaxies based on rotation, isophote shape, and central structure.
The \atlas fast rotators do include roughly half
of the core galaxies, but their rotation-amplitudes are
always at the lower boundary of that subset.
Essentially all core galaxies have \atlas rotation-amplitudes
$\lambda_{R_e/2}\leq0.25,$
while all galaxies with $\lambda_{R_e/2}>0.25$ and figure
eccentricity $>0.2$ lack cores.
Both figure rotation and the central structure of early-type
galaxies should be used together to separate systems that appear to have formed
from ``wet'' versus ``dry'' mergers.

\end{abstract}


\section{The Differences Between Galaxies With and Without Cores}

Luminous elliptical galaxies have long been known to have a
``core'' at their centers, a region interior to which the
steeply-rising stellar surface-brightness profile of the envelope shallows
out to a slowly-rising cusp as $r\rightarrow0$ \citep{l85,k85a, k85b}.
Cores had been predicted to be created when
a binary black hole formed in the merger of two
pre-existing galaxies ejected stars from the
center of the newly formed system \citep{bbr}.
This is still the favored theory for their formation.

With the advent of the {\it Hubble Space Telescope (HST)},
it became possible to observe the central structure of low-luminosity
galaxies that were presumed to have cores too angularly-small
to be detected from the ground.
These galaxies generally lacked cores, however, instead having
steep-cusps in surface brightness as $r\rightarrow0$ \citep{l91,l92}.
As larger samples of early-type galaxies were observed with {\it HST,}
a picture emerged in which the most luminous
elliptical galaxies were confirmed to have cores, but less luminous
galaxies generally did not \citep{crane,k94,f94,l95}.
\citet{l95} called the coreless systems ``power-law'' galaxies,
as they exhibited central light distribution that resembled
steep power-laws as the {\it HST} resolution limit was approached.
The differences between these two kinds of systems extend
to many more properties than their central structure alone,
and in fact define two sub-populations of early-type galaxies
that appear to have different formation histories.

\subsection{Two Families of Elliptical Galaxies}

In parallel with the work on the central structure of early-type
galaxies, observations of the dynamics and isophotal structure of elliptical
galaxies also motivated the recognition that these systems comprise
two distinct groups.  The pioneering work of \citet{defis} showed that
while ``giant'' elliptical rotated slowly, less luminous
systems were consistent with being rotationally supported.
Recent work \citep{e07,c07,a3d3} continues to show that early-type galaxies
are diverse in the degree of organization and amplitude of
their rotation patterns.

Early work with CCD cameras showed that elliptical galaxies
had isophotes that significantly deviated from perfect ellipses,
being either ``boxy'' or `` disky'' \citep{l85b}.
Rich investigations into the isophote shapes of
large samples of elliptical galaxies showed that galaxies
with boxy isophotes correlated with strong radio and X-ray emission,
but slow rotation \citep{bend87, bend88, bend89, nb89}.
Galaxies with disky isophotes showed regular rotation patterns,
but had little radio or X-ray emission.
\citet{n91} unified the link between central structure and the
suite of properties correlated with isophote shape by showing
that galaxies with well-resolved cores had boxy isophotes, while
those with strongly-peaked profiles had disky isophotes.
The ensemble of correlated properties suggested to them that
elliptical galaxies might be divided into two families reflecting
different formation histories.
\citet{tm96} found marked differences between the typical axial
ratios of ellipticals as a function of luminosity, with luminous
ellipticals being nearly round, while faint ellipticals were
significantly flattened, endorsing the segregation of ellipticals
into two families.

\citet{kb96} summarized the picture presented
by these studies of isophote shape and dynamics
of elliptical galaxies, the {\it HST}
surface photometry investigations, and the detailed central
structure analysis of \citet{f97} (then in preparation), to argue
for a explicit bifurcation of elliptical galaxies into two sequences.
One sequence contained the elliptical galaxies with boxy isophotes, cores,
and low rotation-amplitudes.  The second sequence contained elliptical
galaxies with disky isophotes, steep central cusps, and rapid rotation.

\subsection{Core Structure and the Formation of Early Type Galaxies}

\citet{f97} explored the relationships between the physical scale
of the cores in the early-type galaxies,
extending the ground-based core relationships derived by \citet{l85} and
\citet{k85b}, while also testing the link between the form of central
structure and other properties of the galaxies noted by \citet{n91}.
\citet{f97} demonstrated that core and power-law galaxies
were physically distinct subsets of early-type galaxies for more reasons than
their nominal differences in central structure.
The dominant difference
between the two classes is total galaxy luminosity. Most (but not
all) early-type galaxies with $M_V>-20$ are power-laws, while
most (but not all) early-type galaxies with $-22>M_V$ have cores.
In addition to this, \citet{f97} found core galaxies
to rotate slowly, while power-law galaxies rotated rapidly.
Core galaxies also had ``boxy'' isophotes, while power-law
galaxies were ``disky.''
Subsequent work shows that
cores are associated with radio-loud active nuclei \citep{cap05,bal06},
and with strong X-ray emission \citep{pel05},
results that had been anticipated, given the earlier works that linked
radio and X-ray emission to isophote shape.

A crucial finding of \citet{f97} is that rotation and isophote shape
discriminate between core and power-law galaxies of the same
luminosity.  In other words, while the strength of rotation
and isophote shape do correlate with galaxy luminosity, as well,
the correlations are not perfect and independent information
is provided by these parameters.
In particular, over the $-22<M_V<-20$ transition zone in which
Power-law and core galaxies coexist,
properties secondary to galaxy luminosity
still can be used to separate galaxies with the two forms of central structure.
In short, the differences between core and power-law galaxies
are not simply tied to total galaxy luminosity.

The physical differences between the core and power-law galaxies
motivated \citet{f97} to suggest that they were formed by different mechanisms.
Power-law galaxies would be formed
by the mergers of subsystems possessing significant amounts of gas.
During the merger, dissipation would drive gas to the
centers of the galaxies, where it would collapse and form stars
locally, boosting the central stellar density of the merger remnant.
This scenario had been elucidated directly by \citet{mh94}.
Recent detailed simulations of this scenario strongly resemble the
real structure of power-law galaxies \citep{h09}.
Core galaxies, on the other hand, would be formed by the mergers
of systems having little or no gas.  The low rotation of core galaxies
is a direct reflection of their formation in largely stellar,
dissipationless (dry) mergers.
As noted at the start of this introduction, \citet{bbr} hypothesized
that the cores, themselves, would be generated as the hardening
of a black-hole binary formed in the merger would eject stars from
the center of the merged system.  Simulations of this scenario
by \citet{ebi} and subsequent investigators
\citep{mak,q96} supported this hypothesis.
In short, the presence or absence of cores
bears witness to how their surrounding galaxies were formed over all.

The division of early-type galaxies into core and power-law groups
is identical to their division
into two families of elliptical
galaxies by \citet{kb96} and \citet{kfcb}.  This latter work,
continues to find additional differences between the two sets,
extending their dichotomy into differences between the ages
of their stellar populations, the $\alpha$-enhancements of
their chemical abundance patterns, and the \ser indices of their
envelopes.  As is discussed in $\S\ref{sec:core},$ the
distribution of the log-slopes of the inner cusps of
the core and power-law galaxies is essentially bimodal,
making the separation into two groups clear.

The \citet{f97} work was based largely on the relatively small
sample of galaxies observed by \citet{l95}, and was also
limited by the limited availability of long-slit spectroscopic
observations matching the {\it HST} sample.  Subsequent
work has greatly expanded the sample of galaxies with both {\it HST}
imaging and high-quality kinematic information. It is now possible
to re-investigate the relationship between core structure and galaxy
kinematics.

\section{The \atlas Rotation Measurements}

The \atlas project \citep{a3d} used the SAURON IFU spectrograph \citep{sauron}
to obtain high-quality two-dimensional spectroscopy of a large sample
of early-type galaxies.  In brief, \citep{a3d} observed 260 galaxies
brighter than $M_{K_S}=-21.5,$ within 42 Mpc, constrained
by accessible declination and distance from the galactic plane.
\citet{a3d3} (Paper III of the \atlas project) used this material
to derive rotation measures,
\begin{equation}
\lambda_{R_{LIM}}\equiv{\int_0^{R_{LIM}}\int_0^{2\pi} FR|V| dR~d\theta\over
\int_0^{R_{LIM}}\int_0^{2\pi} FR\sqrt{V^2+\sigma^2} dR~d\theta,}
\end{equation}
where $F,$ is the surface brightness distribution
of the galaxy, $R$ is the radius from the center of the galaxy,
$V$ is the stellar rotation field, and $\sigma$ is the stellar velocity
dispersion field.
In a number of systems, \citet{a3d3} were able to integrate out to
$R_{LIM}=R_e,$ the effective radius, but for the great bulk of their sample
they had coverage only out to $R_e/2,$ thus for the remainder
of the paper I use their $\lambda_{Re/2}$ parameters to represent
the degree to which the galaxies are rotating.

\citet{a3d3} use the $\lambda$ measures to build on
their earlier SAURON project \citep{e07,c07}, which introduced
this parameterization and used it to classify early-type galaxies
as either ``fast'' (FR) or ``slow'' (SR) rotators.
Division between the two classes in the
initial SAURON works was set at $\lambda_{R_e}=0.1,$
based on qualitative differences between the rotation curves
of systems on either side of this dividing line.  SR galaxies
had little or no rotation, but also had complex or poorly
organized velocity fields.
They were inferred to be mildly triaxial.
FR galaxies, on the other
hand, had well organized rotation patterns aligned
with the figure axes of the systems, in addition to
the higher amplitude of the rotation over all.
They were inferred to be generally oblate, flattened systems.
In the larger \atlas sample, \citet{a3d3} were better able
to understand the import of projection effects on the observed
rotation amplitudes, and revised the FR/SR boundary to be dependent
on the apparent figure ellipticity of the galaxies:
$\lambda_{Re/2}=0.265\sqrt{\epsilon_{e/2}}.$

\citet{e07}, \citet{c07}, and \citet{a3d3} explored a number of physical
differences between the SR and FR sets and concluded that they likely
had different formation histories.  FR galaxies were inferred to
require gaseous dissipation and star formation during the merger
of their progenitors, while the SR galaxies would reflect the endpoint
of ``dry'' mergers.  As such, the formation of the SR and FR sets
is hypothesized to be the same as that for the ``core'' and ``power-law''
galaxies, respectively. \citet{e07} and \citet{a3d3} did examine
the central structure of galaxies for the subset that had {\it HST}
observations, concluding the SR galaxies generally had cores, and
FR galaxies generally did not. The $\lambda$ dividing
line between the SR and FR classes was set to a very low value, however,
such that the core galaxies were nearly evenly divided between the
FR and SR sets (while the much more numerous power-law galaxies
in their sample still dominate the FR subsample).
Core galaxies that had regular velocity fields
that were well-aligned with their projected figure-axes were typically
assigned to the FR subset.  Again, while a $\lambda$-amplitude
criterion was used to set the boundary between the two classes,
its particular location
was set to select for the {\it qualitative} differences
in the morphologies of the velocity fields.

The high-quality IFU observations of the SAURON and \atlas projects
have been profoundly useful in advancing our understanding
of early-type galaxies.  The motivation of the present work in
revisiting the results of \citet{f97} is to illuminate the link
between the story told by both kinematics and structure.
Because core galaxies fall in both the SR and FR subsets, I am concerned that
their role as diagnostics for the formation of early-type galaxies
risks being over-looked or minimized.
Indeed, this division
has been interpreted in the literature to mean that core and power-law
galaxies have no kinematic differences.
\citet{glass}, for example, summarized the work of
\citet{e07} as showing that there is no clear correspondence between the core
and rotational classes, implicitly negating the conclusions of \citet{f97}.
I argue instead that there is a very deep relationship between the two.

\section{The Detection of Cores in Early-Type Galaxies \label{sec:core}}

The classification of the central structure of early-type
galaxies is provided by the composite sample of \citet{l07a},
which comprises several
{\it HST} studies of the central structure of early-type galaxies
\citep{l95, f97, quil, rav, rest, laine, l05}.
The galaxies represented sample the luminosity range
from dwarf elliptical galaxies to brightest cluster galaxies.
The common thread of the these studies is that they all used
the \citet{l95} ``Nuker law'' to compactly represent the
surface photometry distributions of the galaxies.
There are 63 galaxies in common to the \citet{a3d} and \citet{l07a}
sample, which form the sample studied in this paper.  The list
of galaxies and their parameters are presented in Table \ref{tab:samp}.

In the \citet{l07a} sample, a core is defined to be
the central region of galaxy interior to which the starlight
surface brightness profile takes the form $I(r)\propto r^{-\gamma},$
with $\gamma\leq0.3$ \citep{l95}.  The transition to a core appears as
a ``break'' in the surface brightness profile, a zone over which
the surface brightness profile makes a rapid change in slope
from the steep envelope profile to the shallow cusp interior
to the core, itself.  Power-law galaxies, in contrast, have
steep surface brightness-cusps in their centers, with $\gamma\geq0.5.$
The distribution of the two forms of structure is essentially bimodal
\citep{g96,l07b,kfcb}.
There are a few ``intermediate'' galaxies with $0.3<\gamma<0.5,$
but they are rare.  \citet{l07a} showed that they do not fit on
the core-parameter relations, thus I include the two examples
in the present sample with the power-law galaxies (indeed, their
rotation amplitudes also put them in that set).

Other criteria for identifying cores are possible,
but in practice they agree extremely well with the \citet{l95} formalism.
\citet{k99}, \citet{gra03}, \citet{vcs6}, and \citet{kfcb} for example,
advocated using the centers
of \ser models fitted to the galaxy envelopes as a reference for
determining whether or not a galaxy has a core.  In this schema, a core is
defined to be a central deficit of light with respect to the
\ser model, as opposed to ``excess light,'' which is stellar emission more
centrally concentrated than the \ser model.
In a recent application of this methodology,
\citet{dull} fitted \ser models to the \citet{l05} surface-brightness
profiles, claiming that that Lauer et al.\ misidentified seven
galaxies or 20\%\ of their sample as having cores.
A more objective evaluation is that the ``S\' ersic-reference'' and \citet{l95}
criteria appeared to be in conflict about whether cores were present in
the galaxies in question; they were not misidentified as cores
by \citet{l07a} through incorrect application of their own criteria.
As it happens, however, the concordance between the two
methodologies is actually much better than this.

\begin{figure}
\plotone{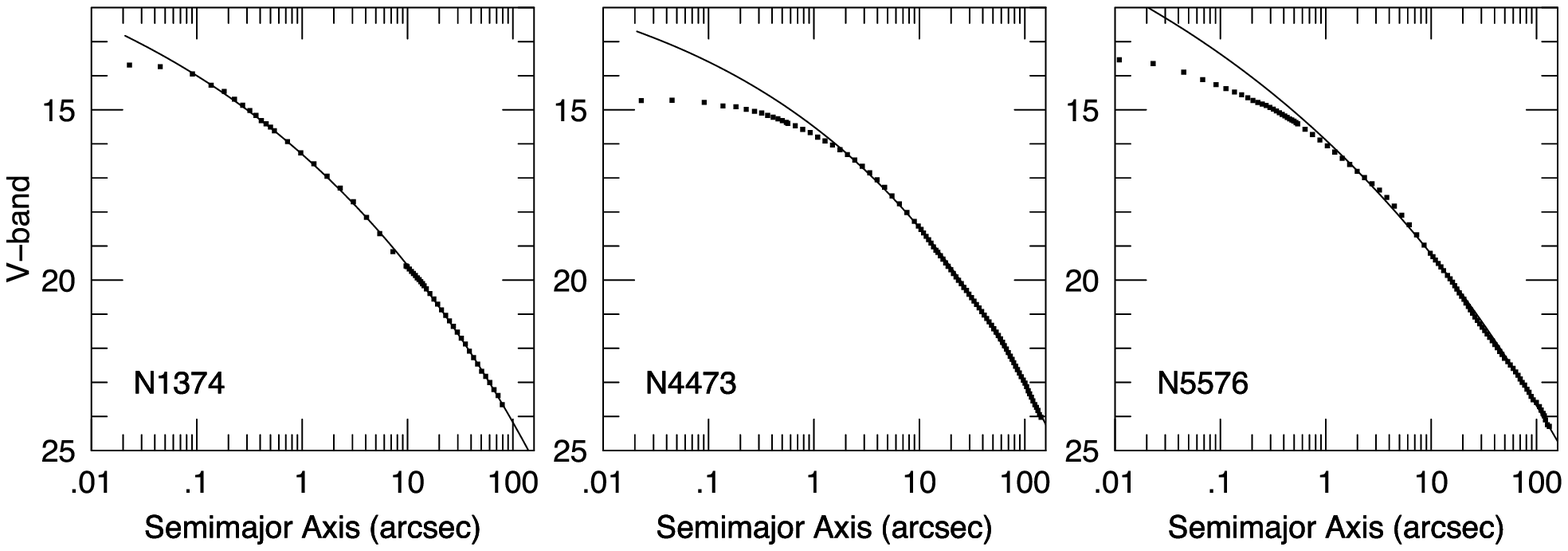}
\caption{\ser model fits are shown for three galaxies claimed by \citet{dull}
{\it not} to have cores.  Central luminosity deficits are clearly
evident in all three galaxies, thus they have cores according to their
own \citet{gra03} criteria, as well as that of \citet{l95}.  The surface
photometry data combines the {\it HST} profiles of \citet{l05} with
ground based data, which provide coverage out to $\sim100''.$
The \citet{dull} fits were done to only the inner $\sim10\%$
of the present profiles, and are thus too limited in radius
to provide accurate representation of the envelopes.}
\label{fig:tricore}
\end{figure}

The seven galaxies in question are NGC 1374, 4458, 4473, 4478, 4486B, 5576,
and 7213.  I show \ser fits to three of these, NGC 1374, 4473, and 5576
in Figure \ref{fig:tricore}, where the surface photometry is a blend of
the high-resolution profiles of \citet{l05}
at small radii with ground-based photometry at larger radii.
The composite profiles extend to radii of $\sim100''.$
\citet{mich} provide the ground-profiles for NGC 4473 and 5576 and
\citet{n1374} provides the profile for NGC 1374.\footnote{I selected
these three galaxies out of the seven because I
had access to complementary ground-based profiles for them.}
As can be seen, the \ser model fits to the envelopes of these galaxies
all show central light deficits or cores in agreement
with the classifications of \citet{l05}.
\citet{dull} in contrast claim that these galaxies have no central deficits,
but their fits were done only over the inner regions
of the galaxies that were sampled by the {\it HST} photometry,
which covers only the inner $\sim10\%$ of the present profiles.
The \citet{dull} \ser fits are thus too limited in radius
to accurately characterize the envelopes.

Of the remaining galaxies, NGC 4486B actually
shows a central deficit with respect to the \citet{dull} \ser model,
but they reject a core classification,
claiming that the plateau in the inner light profile
is not due to a relative deficit of stars.
They do not justify this contradictory statement.
Additional guidance comes from the \ser fits to
profiles derived from large-format imagers provided by \citet{kfcb}.
These investigators emphasize the need to obtain accurate surface photometry
at large radii to enable the correct characterization of the envelope.
\citet{kfcb} actually do find excess central light in N4486B, as
well as for NGC 4458 and N4478.
They also provide two \ser models for NGC 4473,
one of which shows a central light deficit, while the other is
fitted to a more restricted envelope domain and shows central excess light.
\citet{kfcb} note that NGC 4473 appears to have a core on morphological
grounds, but classify it instead as a galaxy with excess light based
on considerations of its central kinematics,
which indicate the existence of a central counter-rotating
stellar disk \citep{sauron3}.
Regardless, the physical scale of the NGC 4473 core is normal for its
luminosity \citep{l07a}.
In the case of NGC 4458, the \citet{kfcb}
\ser fit falls below the surface brightness
profile for $r<300$ pc, implying that the central component
is an extended system well over an order of magnitude larger than typical
nuclear star clusters \citep[see the discussion in][]{l07b}.
The \citet{l05} core determination is for this inner component,
not the envelope component described by the \ser model.

I conclude that the concordance
between the very different \citet{l95} and \citet{gra03}
criteria for the identification of cores is at least at the $\sim90\%$\ level,
a conclusion already reached by \citet{kfcb}.  However, it is clear
that the use of \ser models to reliably identify cores requires having
profiles of large radial extent.
Oddly, one relies on the precise form of the galaxy on
scales of several kiloparsecs to evaluate the nature of the central structure
on scales of a few hundred parsecs or less.
Their use may also require subjective
choices on the domain over which the models are fitted.
The criteria of \citet{l95}, in contrast, are
local to the center of the galaxies.
In the next section
I also show that they are more likely to select galaxies that
are rotationally consistent with having cores when the
\ser models indicate excess central light.

\section{Core Galaxies Don't Rotate Very Much, Unlike Power-Law Galaxies}

Figure \ref{fig:mv_rot} plots $\lambda_{Re/2}$ as a function of
total galaxy luminosity, $M_V,$ with the symbols encoding whether
or not the galaxy has core or steep power-law cusp.
As can be seen, all core galaxies have $\lambda_{Re/2}<0.32,$
a limit that can be decreased to 0.25, if NGC 3640 is excluded.
The median $\lambda_{Re/2}$ for core galaxies is only 0.09,
compared to 0.39 for power-law galaxies.
The strong segregation of the core and power-law galaxies
confirms the conclusion of \citet{f97} that the two forms of
central structure correspond to different levels of rotation
in the larger bodies of the galaxies.

There are two additional points worth noting. First, as found in \citet{f97},
while on average faint galaxies rotate rapidly, and luminous galaxies do not,
over the luminosity interval over which core and power-law
galaxies coexist, the strength of rotation remains a strong
discriminant between the two types, as is clearly evident
in Figure \ref{fig:mv_rot} for galaxies with $-22<M_V<-21.$

The second point is that faint core galaxies with $M_V>-20$ still
have low levels of rotation consistent with the luminous
core galaxies, rather than power-law galaxies of the same luminosity.
These are rare systems with particularly compact cores, thus
their connection to core galaxies with $M_V<-21$ might have been
questioned.  The two core galaxies in the present
sample with $M_V>-20$ are NGC 4458 and 4478.  It is notable
that both galaxies are cases in which the classification
of core structure by the \citet{l95} and \citet{gra03} criteria
disagreed.  \citet{a3d3} indeed flagged NGC 4458 as an anomalous
case of a galaxy with a large luminosity excess, but yet that
falls into their SR class.

The identification of NGC 4458 and 4478 as core galaxies
by the \citet{l95} criteria, as opposed to their classification
as galaxies with central light excesses when referenced to \ser models,
served as the best predictor of their rotation measures.
Of the other galaxies discussed in the previous section, NGC 5576
is classified as an SR by \citet{a3d3}, and NGC 4473 is near the
top of the $\lambda_{Re/2}$ range occupied by core galaxies, but
also is highly elongated (see the discussion below); \atlas has no
data on NGC 1374, 4486B, or 7213.

While power-law galaxies do rotate faster than core galaxies,
on average, the Figure \ref{fig:mv_rot}
does show that there are a few power-law
galaxies that fall among the core galaxies.
This is almost certainly due to projection
effects.  If the rotation axis of a galaxy falls close to the
line of sight, such that the galaxy is viewed largely ``face on,'' then
the amplitude of the rotation observed will be greatly reduced.

To explore this possibility, I have also encoded the ellipticities
of the power-law galaxies in Figure \ref{fig:mv_rot},
using a value of $\epsilon_{e/2}=0.2$ to separate them into two subsets.
All of the power-law galaxies with $\lambda_{Re/2}\leq0.25$ in fact also
have low ellipticities, as compared to the generally-high
ellipticities of these galaxies.  The suggestion is
that these galaxies indeed are face-on to the line of sight.
\citet{l05} also showed that low-ellipticity power-law galaxies
were likely to be preferentially face-on.  Their Figure 6,
reproduced here as Figure \ref{fig:l05_ell}, shows that
inner stellar disks are evident in power-law galaxies only
when their inner $\epsilon>0.25$ --- the implication is that
all power-law galaxies have disks, a point made initially by
\citet{f94}, but are not seen below this dividing line because
the galaxies are inclined with respect to the line-of-sight.

The relationship between the projected ellipticity of the sample
galaxies and their rotation measures is explored further
in Figure \ref{fig:ell_rot}, which directly compares both parameters,
a plot that was heavily emphasized by \citet{a3d3} in their own
analysis of the distribution of $\lambda_{Re/2}$ as a
function of galaxy properties.
Power-law galaxies clearly have preferentially higher ellipticity,
while core galaxies are preferentially rounder, in general
agreement with the results of \citet{tm96}, who explored the relationship
between axis ratios and luminosity.
At the same time, the handful of core galaxies with high ellipticity
still have small $\lambda_{Re/2}.$

The solid line plotted in Figure \ref{fig:ell_rot} shows the
dividing line, $\lambda_{Re/2}=0.265\sqrt{\epsilon_{e/2}},$
adopted by \citet{a3d3} to discriminate between their SR and FR classes.
While this separation does imply that nearly all SR galaxies are core galaxies,
the converse is not true, as roughly half of the core galaxies
also fall into the FR class.  The dotted line in Figure \ref{fig:ell_rot}
shows an alternative dividing line of $\lambda_{Re/2}=0.25.$  This
frees the FR class of any ``contamination'' of core galaxies, and
leaves only a relatively small number of presumably face-on power-law
galaxies in the SR class.

An interesting question is whether or not the cores in the nominal
\citet{a3d3} FR versus SR classes reflect important physical differences
between the two sets, apart from the regularity and amplitude
of their rotational fields.
In the present sample, the SR core galaxies correspond
to those with $\lambda_{Re/2}\leq0.10,$ or 15 galaxies.  It does
appear that these galaxies may be preferentially more luminous.  The SR
core galaxies have a median $M_V=-22.0,$ which can be compared
to the median $M_V=-21.5$ of the 10 FR core galaxies.
Notably, of the nine core galaxies
with $M_V<-22,$ only one of them, NGC 4649, is a FR galaxy.

An obvious hypothesis is that if cores reflect the endpoint
of dry mergers, then the SR galaxies might be systems in which
more than one dry merging event has built up the systems, resulting
in more thorough erasure of the originally regular rotation
patterns of the presumed power-law galaxy progenitors.
There is no signature of this in the core
structure of the systems, however.  If I naively assume that
more generations of merging might produce relatively
larger cores in the SR versus FR core galaxies, I might expect that
residuals in the the $M_V-r_\gamma$ relation of \citet{l07a},
which relates the size of the cores, $r_\gamma,$ to galaxy luminosity,
would be correlated with $\lambda_{Re/2}.$  The mean relation
in \citet{l07a} is $r_\gamma\propto L_V^{1.9\pm0.03}.$  Residuals
about this relation are plotted in Figure \ref{fig:res_core} as
a function of $\lambda_{Re/2}.$  There is no suggestion of any
trend in this graph, thus no indication that there are any interesting
differences in at least the core structure of the FR versus SR
core galaxies.

The slope of the $M_V-r_\gamma$
relation, itself, is unremarkable.  In terms of {\it mass} of the
stars associated with the core and the total stellar mass of the galaxy,
the relations in \citet{l07a} imply $m_\gamma\propto M_*^{1.1\pm0.1},$
where $m_\gamma$ is the ``core mass,'' and $M_*$ is the stellar
mass of the galaxy, itself.  The essentially linear proportionality,
means that the cores are not growing faster than their surrounding
galaxies over the luminosity interval in which FR core galaxies
might be converted to SR core galaxies.

\section{Core Morphology and Kinematics Working in Harmony}

\citet{f97} divided early-type galaxies into core and power-law galaxies,
concluding that the two types had different formation histories.
\citet{kb96} and \citet{kfcb} divided early-type galaxies into two
classes, concluding that the two types had different formation histories.
\citet{e07} divided early-type galaxies into SR and FR galaxies,
concluding that the two types had different formation histories.
The description of the SR and FR formation hypotheses sounds more
or less the same as those for core and power-law galaxies.
My interpretation is that the observations of \citet{e07} and \citet{a3d3}
not only confirm the conclusion that
cores and power-laws have markedly different dynamics, but also show that
central structure serves
as a sharp way to sort out the two formation families
of early-type galaxies  common the all of these investigations.
I summarize my reasoning as follows:

\begin{itemize}

\item Core galaxies have median $\lambda_{Re/2}=0.09;$ 
Power-law galaxies have median $\lambda_{Re/2}=0.39.$
There is an unambiguous difference in their mean dynamical properties.

\item All core galaxies, except one, have $\lambda_{Re/2}\leq0.25.$
All power-law galaxies with $\epsilon_{e/2}>0.2$ have $\lambda_{Re/2}>0.25.$
The two sets are essentially completely disjoint.

\item All power-law galaxies with $\epsilon_{e/2}\leq0.2,$ but for one,
have $\lambda_{Re/2}\leq0.25.$  Power-laws with inner ellipticity
$>0.25$ mostly have inner stellar disks, rounder than 0.25 never do.
Slowly rotating power-law galaxies are those simply viewed
from an unfavorable angle.  Indeed such systems must be observed
by chance in large samples, even if all power-law galaxies
are intrinsically fast rotators.

\item Over the interval $-22<M_V<-21,$ in which core and power-law
galaxies coexist, the two subsets remain sharply segregated by
their $\lambda_{Re/2}$ values.  The two core galaxies with
$M_V>-20$ are segregated from power-law galaxies of the same
luminosity by their low $\lambda_{Re/2}$ values.  Power-law galaxies
are typically much fainter than core galaxies, but at all luminosities
where the two classes coexist they can be sharply separated by
rotation amplitude and projected ellipticity.  The differences
between core and power-law galaxies are not a trivial consequence
of their differing average luminosities.

\end{itemize}

Over all, the observations are consistent with the hypothesis that
core and power-law galaxies have completely disjoint dynamical properties.
One can clearly enhance the segregation of core and power-law galaxies
in the \atlas sample by simply boosting the SR and FR dividing line to
$\lambda_{Re/2}=0.25,$ and perhaps requiring that the FR galaxies also
have $\epsilon_{e/2}>0.2$, as well.\footnote{\citet{kb12}
have also noted that moving the $\lambda_{Re/2}$ boundary upwards
would lead to better separation of core and power-law galaxies.}
This, of course, comes with the risk that more truly fast-rotating
galaxies will fall into the SR class by virtue of unfavorable projection.
A potential solution is to
obtain central structure measurements on all galaxies
in the \atlas sample with $\epsilon_{e/2}<0.2$ that presently lack them,
such that a sample complete in both structure and kinematics
can be constructed.  While rotation measures depend on the
inclination of the galaxies to the line-of-sight, the structure
measures are independent of viewing angle and
could be used to sort out which rotation class the rounder galaxies
with low $\lambda_{Re/2}$ are most likely to belong to.

\begin{figure}
\plotone{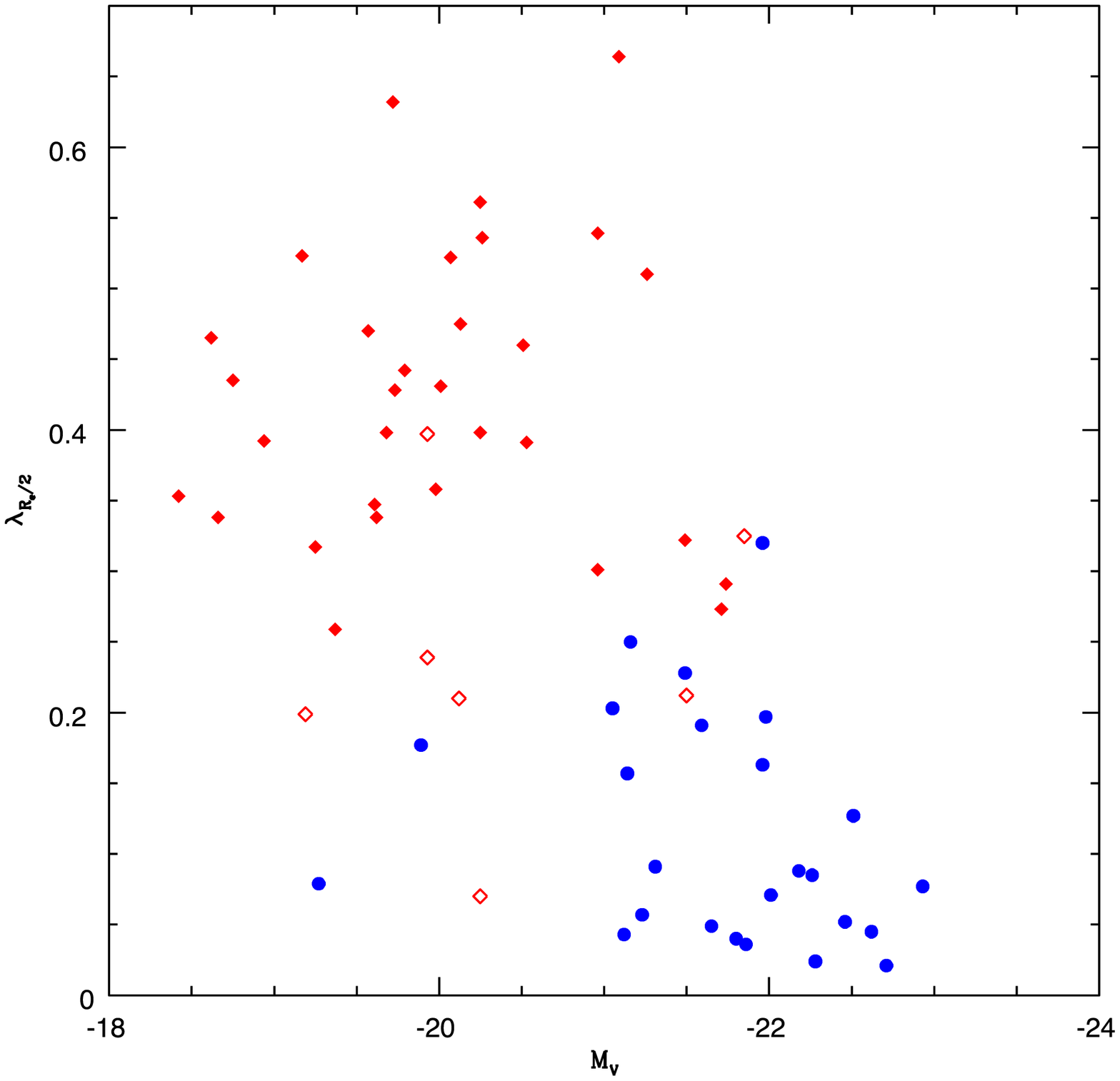}
\caption{Galaxy rotation as measured by the \citet{a3d3} $\lambda_{Re/2}$
parameter is plotted as a function of total luminosity
for the galaxies in common between the \citet{a3d3} and \citet{l07a}
samples.  Core galaxies are plotted as round-blue symbols, while power-laws
are plotted as red-diamonds.  Power-law galaxies with $\epsilon_{e/2}\leq0.2,$
which are likely to have low rotation due to projection effects,
are further plotted as open diamonds.  A clear separation
of core and power-galaxies is seen, even in the luminosity
range over which the two types coexist.}
\label{fig:mv_rot}
\end{figure}

\begin{figure}
\plotone{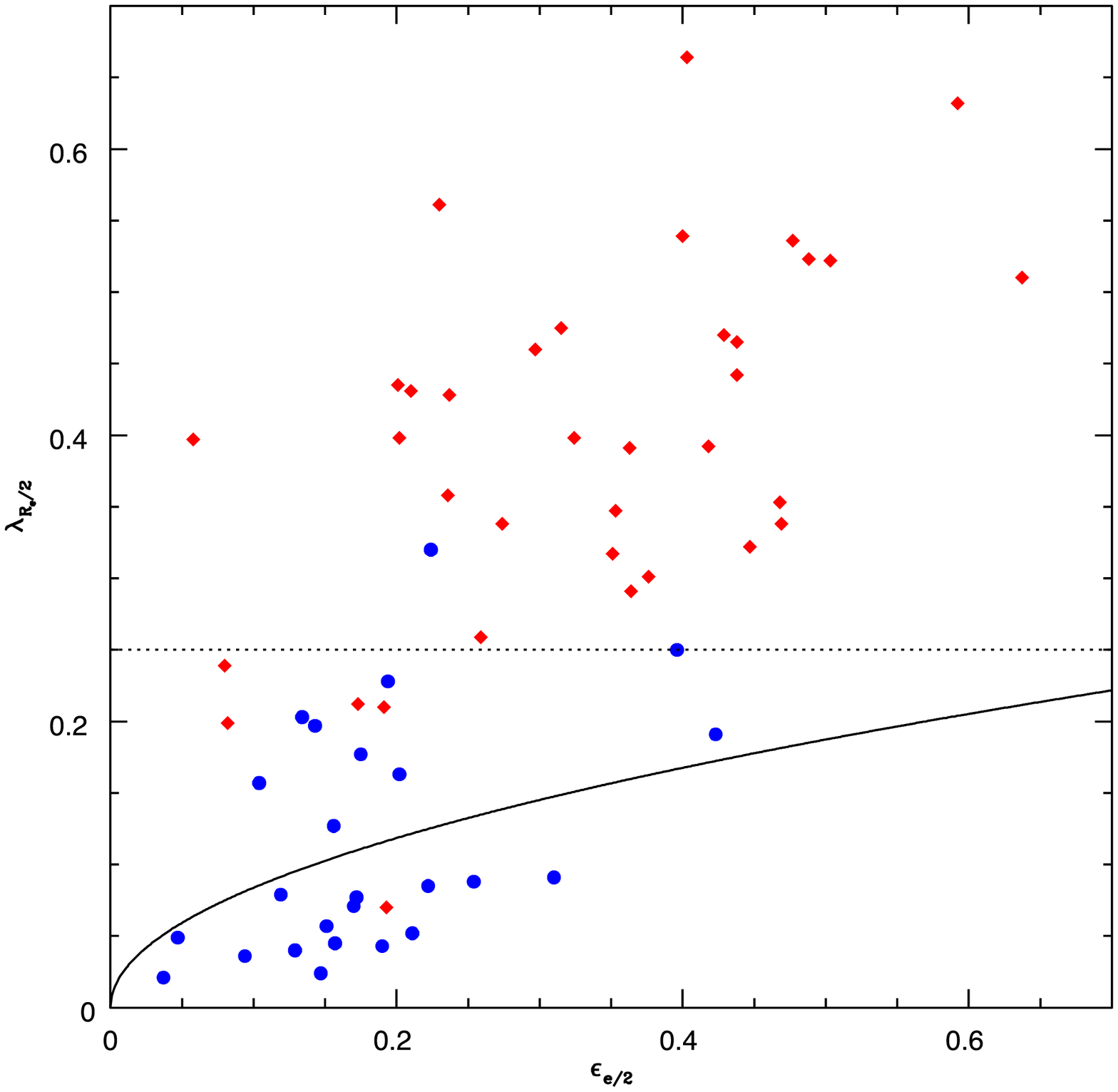}
\caption{Galaxy rotation as measured by the \citet{a3d3} $\lambda_{Re/2}$
parameter is plotted as a function of average ellipticity
for the galaxies in common between the \citet{a3d3} and \citet{l07a}
samples.  Core galaxies are plotted as round-blue symbols, while power-laws
are plotted as red-diamonds.  The solid line is the separation
between FR and SR galaxies suggested by \citet{a3d3}.
The dotted line at $\lambda_{Re/2}=0.25$ neatly
separates the core and power-law galaxies into different rotation
classes, leaving the core set only contaminated by face-on power-law galaxies.} 
\label{fig:ell_rot}
\end{figure}

\begin{figure}
\plotone{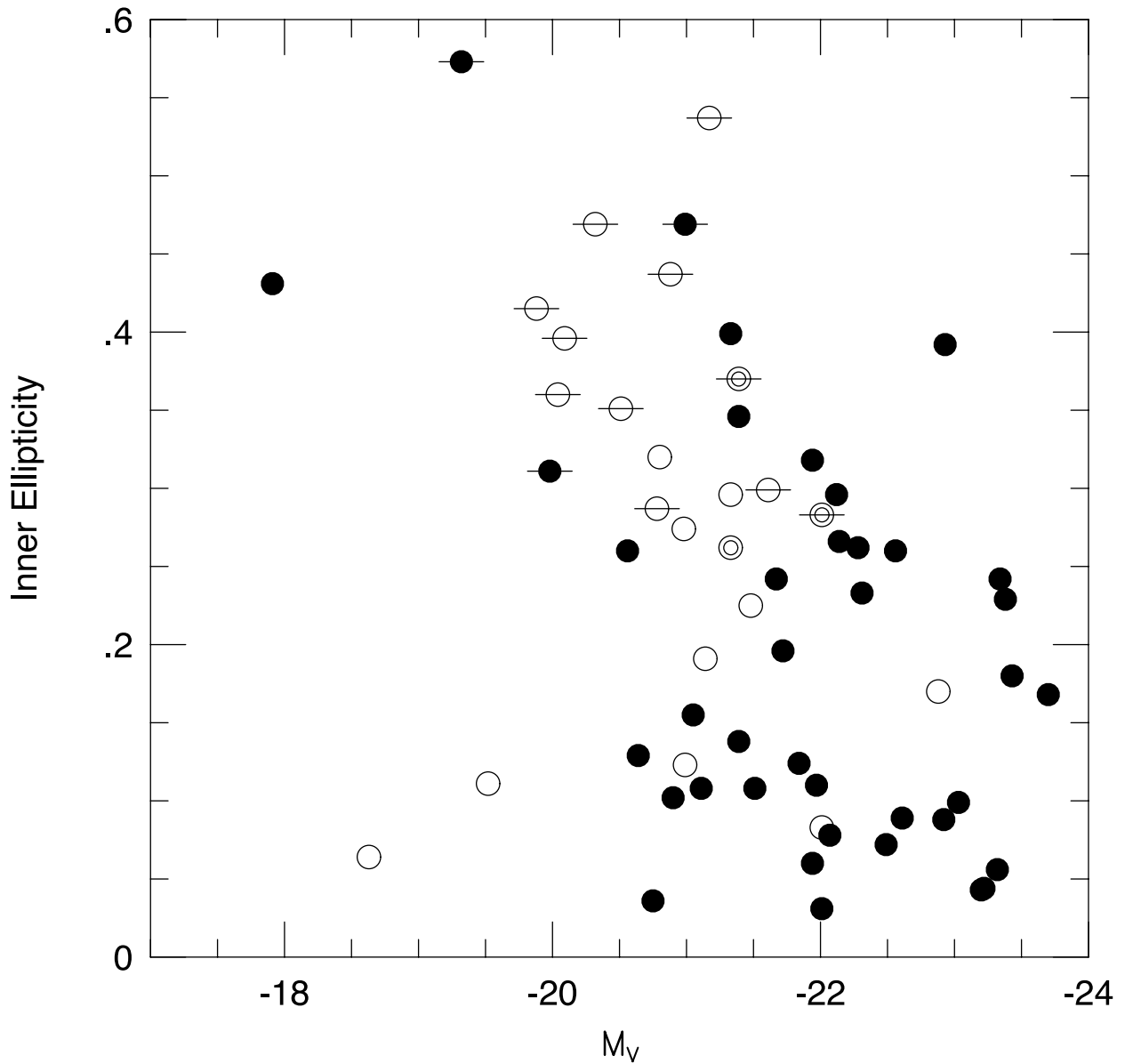}
\caption{Inner luminosity-weighted isophote ellipticity is
plotted as a function of total galaxy luminosity, as shown
in Figure 6 of \citet{l05}.
Solid symbols are core
galaxies, open symbols are power-law galaxies, and intermediate galaxies
are plotted as double open circles.  Galaxies with {\it inner} stellar
disks are indicated with horizontal lines.
Nearly all power-law
galaxies with $\epsilon\geq0.3$ have inner disks, implying that they
are present in the rounder power-law galaxies, but are not seen due to
unfavorable viewing angles.  Disks are visible in flattened core
and intermediate galaxies fainter than $M_V\approx-21,$ suggesting
that these are transitional objects.  Disks are not seen in bright
core galaxies.}
\label{fig:l05_ell}
\end{figure}

\begin{figure}
\plotone{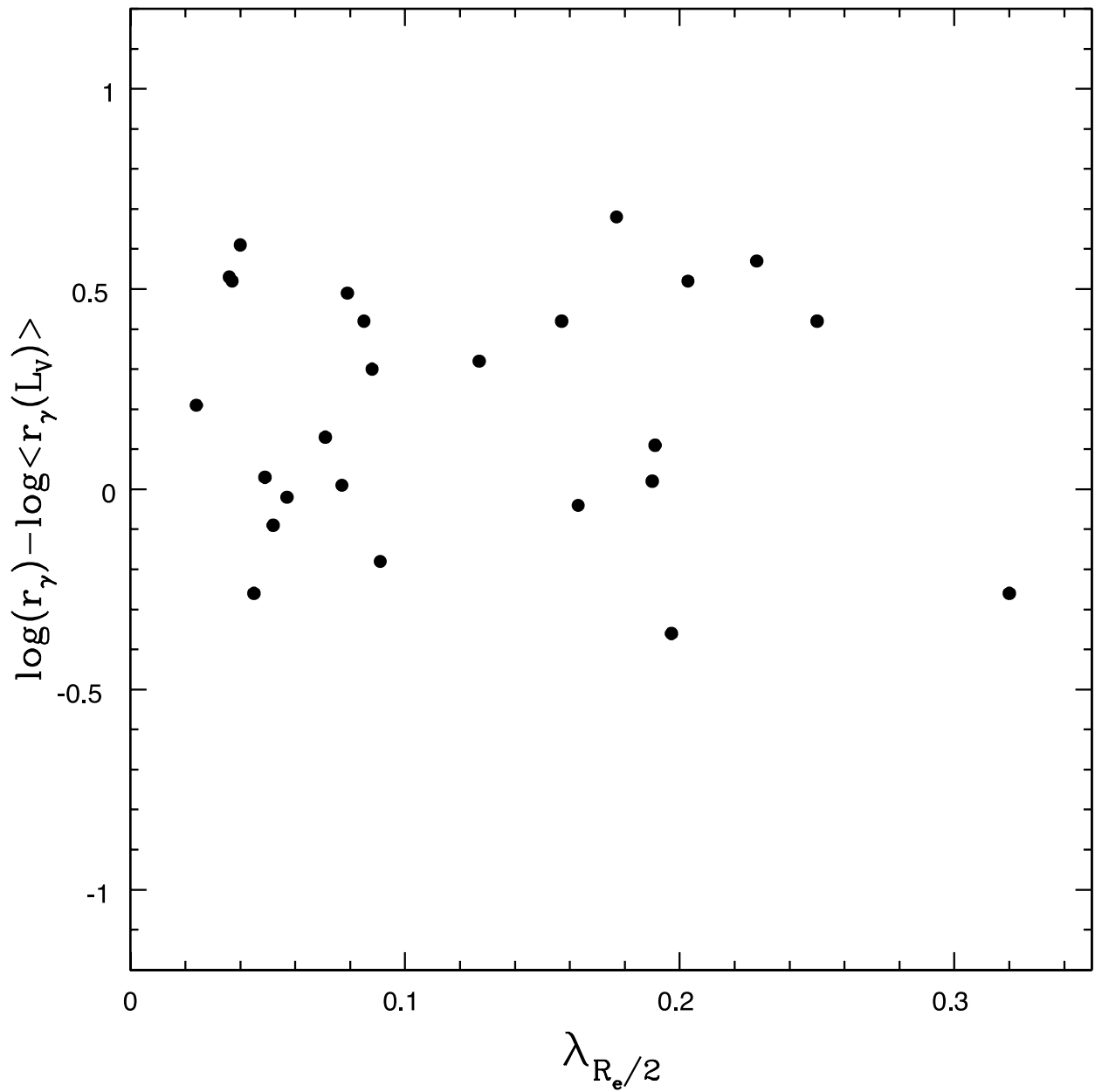}
\caption{Residuals in the core physical scale, $r_\gamma,$ in the
core galaxies from the mean relation between $r_\gamma$ and
galaxy luminosity \citep[equation 14 in][]{l07a} are plotted
as a function of $\lambda_{Re/2}$ from \citet{a3d3}. No trend
is evident, so there is no indication that cores in SR core galaxies,
that is those with $\lambda_{Re/2}\leq0.1,$ are preferentially
larger than those in FR core galaxies.}
\label{fig:res_core}
\end{figure}

The obvious question is does this matter.
If one focusses on the properties of galaxies on an individual basis
and avoids dividing-lines,
as can be done with the parameter plots in Figures \ref{fig:mv_rot}
or \ref{fig:ell_rot}, the answer is no.
The sense of \citet{kb96}, \citet{f97}, and \citet{e07}, however, that it
does make sense to consider discreet families of early-type galaxies.
Sorting objects into classes risks losing important information,
but not doing so when they really do appear to correspond
to {\it qualitatively} different origins risks missing the big picture.
The problem then is to take care when advancing statements of the sort,
``Fast-rotators are this, while slow-rotators are that.''
For example, while one could say, ``Core galaxies are equally divided
between the FR and SR classes" and be technically correct,
such a statement completely misses the critical details of
picture presented by Figures \ref{fig:mv_rot} and \ref{fig:ell_rot}.

The unique information captured
by the SAURON and \atlas\ projects is the form of the velocity fields.
On that side of town, the concern is understanding the
physics that turns less-luminous systems with regular velocity
fields into the complex and minimally-rotating SR velocity fields.
On the side of town that I come from, on the concern has been understanding
the physics that turns less-luminous systems that have no cores
into luminous galaxies that do.  The contested ground is that occupied
by galaxies that have cores with regular rotation fields,
but of considerably lower-amplitude than those in FR systems lacking cores.
Obviously, the creation of cores and the erasure of regular
rotation patterns, as more luminous early-type galaxies
are built from mergers, are not exactly synchronized.  Both processes
happen over a range of luminosity, largely but not completely
bounded by $-20>M_V>-22.$

In trying to understand what happens over this range,
I come back to the galaxies with $M_V>-20$ that I argue have cores.
It is true that I cannot know if, say NGC 4458, and 4478
``really'' have cores, in the
sense that their central structure reflects the same ``core-scouring''
mechanisms that is hypothesized to set the form of the highly-luminous SR core
galaxies with $M_V<-22$ at the end of the line.
But at the same time, no-one knows what the full luminosity range
of systems that might be generated by ``core-scouring'' look like,
nor what the kinematics of those objects might be,
particularly if the initial mergers are of unequal mass,
and some amount of gas (as in ``damp'' mergers) is present in the first steps.
If there is a road by which high-luminosity elliptical galaxies
are built from the mergers of rapidly-rotating low-luminosity
elliptical galaxies, finding its start may be best done by
using morphology and kinematics together to find the first galaxies
along its path.

\acknowledgments

I thank Eric Emsellem, Doug Richstone, and Tim de Zeeuw
for useful conversations.  I also appreciate the thorough
and prompt review by the referee that bolstered the arguments
presented in the paper.

\clearpage

\begin{deluxetable}{cccccccc}
\tablecolumns{8}
\tablewidth{0pt}
\tablecaption{Rotation and Core Classifications}
\tablehead{\colhead{}&\colhead{}&\colhead{}&\colhead{D}
&\colhead{}&\colhead{}&\colhead{} &\colhead{} \\
\colhead{Galaxy}&\colhead{Morph}&\colhead{$M_V$}&\colhead{(Mpc)}
&\colhead{P}&\colhead{$\lambda_{R_e/2}$}&\colhead{$\epsilon_{e/2}$}
&\colhead{Ref}}
\startdata
N0474&S0&$-$20.12& 29&$\backslash$&0.21&0.19&6\\
N0524&S0+&$-$21.85& 25&$\wedge$&0.33&0.03&4\\
N0821&E&$-$21.71& 25&$\wedge$&0.27&0.39&1\\
N1023&S0-&$-$20.53& 12&$\backslash$&0.39&0.36&1\\
N2549&S0&$-$19.17& 13&$\backslash$&0.52&0.49&2\\
N2592&E&$-$20.01& 27&$\backslash$&0.43&0.21&2\\
N2685&S0+&$-$19.72& 14&$\backslash$&0.63&0.59&6\\
N2699&E&$-$20.25& 28&$\backslash$&0.40&0.20&2\\
N2778&E&$-$18.75& 24&$\backslash$&0.43&0.20&1\\
N2950&S0&$-$19.73& 15&$\backslash$&0.43&0.24&2\\
N2974&E&$-$21.09& 22&$\backslash$&0.66&0.40&1\\
N3193&E&$-$21.98& 36&$\cap$&0.20&0.14&2\\
N3377&E&$-$20.07& 11&$\backslash$&0.52&0.50&1\\
N3379&E&$-$21.14& 11&$\cap$&0.16&0.10&1\\
N3384&S0-&$-$19.93& 11&$\backslash$&0.40&0.06&1\\
N3414&S0&$-$20.25& 26&$\backslash$&0.07&0.19&2\\
N3595&E&$-$20.96& 35&$\backslash$&0.30&0.38&2\\
N3599&S0&$-$19.93& 23&$\backslash$&0.24&0.08&3\\
N3605&E&$-$19.61& 23&$\backslash$&0.35&0.35&3\\
N3607&S0&$-$21.49& 21&$\cap$&0.23&0.19&1\\
N3608&E&$-$21.12& 23&$\cap$&0.04&0.19&1\\
N3610&E&$-$20.96& 22&$\backslash$&0.54&0.40&1\\
N3613&E&$-$21.59& 30&$\cap$&0.19&0.42&2\\
N3640&E&$-$21.96& 28&$\cap$&0.32&0.22&1\\
N3945&S0+&$-$20.25& 19&$\backslash$&0.56&0.23&1\\
N4026&S0&$-$19.79& 15&$\backslash$&0.44&0.44&1\\
N4143&S0&$-$19.68& 15&$\backslash$&0.40&0.32&6\\
N4150&S0&$-$18.66& 14&$\backslash$&0.34&0.27&4\\
N4168&E&$-$21.80& 37&$\cap$&0.04&0.13&2\\
N4261&E&$-$22.26& 33&$\cap$&0.09&0.22&4\\
N4278&E&$-$21.05& 16&$\cap$&0.20&0.13&1\\
N4365&E&$-$22.18& 21&$\cap$&0.09&0.25&1\\
N4374&E&$-$22.28& 17&$\cap$&0.02&0.15&4\\
N4382&S0+&$-$21.96& 17&$\cap$&0.16&0.20&1\\
N4387&E&$-$19.25& 17&$\backslash$&0.32&0.35&3\\
N4406&E&$-$22.46& 17&$\cap$&0.05&0.21&1\\
N4417&S0&$-$18.94& 17&$\backslash$&0.39&0.42&6\\
N4434&E&$-$19.19& 17&$\backslash$&0.20&0.08&3\\
N4458&E&$-$19.27& 17&$\cap$&0.08&0.12&1\\
N4472&E&$-$22.93& 17&$\cap$&0.08&0.17&1\\
N4473&E&$-$21.16& 17&$\cap$&0.25&0.40&1\\
N4474&S0&$-$18.42& 21&$\backslash$&0.35&0.47&2\\
N4478&E&$-$19.89& 17&$\cap$&0.18&0.17&1\\
N4486&E&$-$22.71& 17&$\cap$&0.02&0.04&3\\
N4494&E&$-$21.50& 17&$\backslash$&0.21&0.17&1\\
N4503&S0-&$-$19.57& 17&$\backslash$&0.47&0.43&2\\
N4551&E&$-$19.37& 17&$\backslash$&0.26&0.26&3\\
N4552&E&$-$21.65& 17&$\cap$&0.05&0.05&1\\
N4564&E&$-$20.26& 17&$\backslash$&0.54&0.48&2\\
N4621&E&$-$21.74& 17&$\backslash$&0.29&0.36&1\\
N4636&E&$-$21.86& 17&$\cap$&0.04&0.09&4\\
N4649&E&$-$22.51& 17&$\cap$&0.13&0.16&1\\
N4660&E&$-$20.13& 17&$\backslash$&0.47&0.32&1\\
N4697&E&$-$21.49& 13&$\backslash$&0.32&0.45&3\\
N5198&E&$-$21.23& 38&$\cap$&0.06&0.15&2\\
N5308&S0-&$-$21.26& 33&$\backslash$&0.51&0.64&2\\
N5557&E&$-$22.62& 52&$\cap$&0.04&0.16&1\\
N5576&E&$-$21.31& 27&$\cap$&0.09&0.31&1\\
N5813&E&$-$22.01& 28&$\cap$&0.07&0.17&1\\
N5838&S0-&$-$20.51& 22&$\backslash$&0.46&0.30&6\\
N5845&E&$-$19.98& 28&$\backslash$&0.36&0.24&4\\
N7332&S0&$-$19.62& 24&$\backslash$&0.34&0.47&3\\
N7457&S0-&$-$18.62& 14&$\backslash$&0.47&0.44&1\\
\enddata
\label{tab:samp}
\tablecomments{
Morphological classifications are from the RC3 \citep{rc3}.
Distances and total luminosity assume $H_0=70$ km s$^{-1}$ Mpc$^{-1}.$
The profile type, P, is $\backslash=$ power-law,
$\wedge=$ intermediate form, and $\cap=$ core.  The reference column refers
to the origin of central structural
parameters for the given galaxy as follows: 1) \citet{l05}; 2) \citet{rest};
3) \citet{l95} or \citet{f97}; 4) \citet{quil}; and 5) \citet{rav}.}
\end{deluxetable}

\end{document}